# Asymmetry interactions of electrons and positrons in strong pulsed laser fields

Stanislav Starodub[1], Sergei Roshchupkin[2] and Viktor Dubov[2]

[1]Institute of Applied Physics, National Academy of Sciences of Ukraine,

58 Petropavlovskaya Str., Sumy 40000, Ukraine

[2]Department of Theoretical Physics, Peter the Great St. Petersburg Polytechnic University,

195251, St-Petersburg, Russian Federation, Russia

E-mail: roshchupkin_s@spbstu.ru

**Abstract.** The interaction of two electrons and two positrons in the field of two strong pulsed laser waves outside the framework of the dipole approximation (with an accuracy of relativistic corrections $v/c$) has been studied theoretically. The possibility of significant asymmetry in the interactions of particles and antiparticles in the external laser fields is shown. This asymmetry of interactions essentially depends on the magnitude and sign of the phase shifts of the waves relative to each other, and also on the intensities of the waves. The ranges of phase shifts and field intensities in which the interaction of electrons and positrons have qualitatively different character from anomalous repulsion to attraction are determined.

**Key words:** nonrelativistic electrons and positrons, strong femtosecond laser pulsed fields, effective interaction, phase shift.

PACS: 41.20.Jb, 41.75.Fr

**1 Introduction**
There are many works devoted to research of interaction of electrons in the presence of an electromagnetic field (see, the works [1-9]). The possibility of electron attraction in the presence of a plane electromagnetic wave was firstly assumed by Oleinik [4]. However, the theoretical proof of the attraction possibility was given by Kazantsev and Sokolov for interaction of classical relativistic electrons in the field of a plane wave [5]. We also note the paper [6]. It is very important to point out, that attraction of classical electrons in the field of a plane monochromatic electromagnetic wave is possible only for particles with relativistic energies. In the authors works (see, review [3], articles [7-9]) the possibility of attraction of nonrelativistic electrons (identically charged ions) in the pulsed laser field was shown. Thus, in the review [3] the following processes were discussed: interaction of electrons (light ions) in the pulsed field of a single laser wave; interaction of nonrelativistic electrons in the pulsed field of two counter-propagating laser waves moving perpendicularly to the initial direction of electrons motion; the interaction of nonrelativistic light ions moving almost parallel to each other in the propagation direction of the pulsed field of two counter-propagating laser waves moving in parallel direction to ions; interaction of two nonrelativistic heavy nuclei (uranium 235), moving towards each other perpendicularly to the propagation direction of two counter-propagating laser waves. The effective force of interaction of two hydrogen atoms (after their ionization) in the pulsed field of two counter-propagating laser waves was considered in [7]. Influence of pulsed field of two co-propagating laser waves on the effective force of interaction of two electrons and two identically charged heavy nuclei was studied in [8]. The main attention is focused on the study of the influence of phase shifts of the pulse peak of the second wave relatively to the first on the effective force of particles interaction. The phase shift allows to increase duration of electron's confinement at a certain averaged effective distance by 1,5 time in comparison with the case of one and two counter-propagating pulsed laser waves. Interaction of two classical nonrelativistic electrons in the strong pulsed laser field of two light mutually perpendicular waves, when the maxima laser pulses coincide, was studied in [9]. It is shown that the effective force of electron interaction becoming the attraction force or anomalous repulsion force after approach of electrons to the minimum distance.

In this paper, in contrast to the previous papers, the effective interaction of two electrons and two positrons in the field of mutually perpendicular strong laser waves is studied with allowance for their phase shifts of the pulses. The ranges of intensities and phase shifts of waves are determined for which the effective interaction of two particles (electrons) and two antiparticles (positrons) is essentially asymmetric.

## 2. Equations of particles interaction in pulsed field of two laser waves

We study interaction of two nonrelativistic electrons (positrons) moving towards each other along the axis $x$ in a field of two linearly polarized pulsed electromagnetic waves. Waves propagate perpendicularly to each other. The first wave propagates along the axis $z$, the second wave propagates along the axis $x$ (see figure 1).

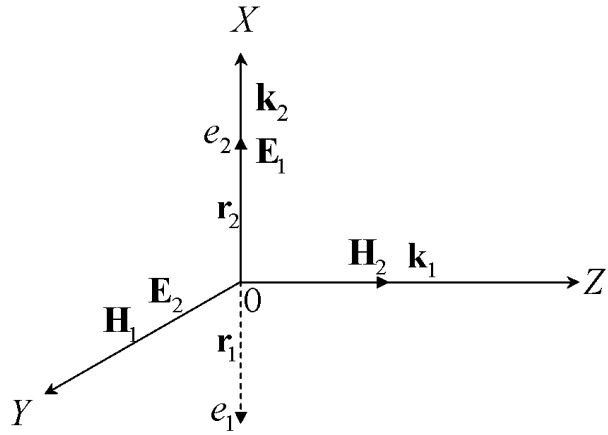

**Figure 1.** Interaction kinematics of two classical electrons (positrons) in the field of two light mutually perpendicular waves.

We assume the strength of electric and magnetic field in following form:

$$\boldsymbol{E}(t, z_j, x_j) = \boldsymbol{E}_1(t, z_j) + \boldsymbol{E}_2(t, x_j), \tag{1}$$

$$\boldsymbol{E}_1(t, z_j) = E_{01} \cdot \exp\left[-\left(\frac{\varphi_{1j} - \delta\tau_1}{\omega_1 t_1}\right)^2\right] \cos\varphi_{1j} \cdot \boldsymbol{e}_x, \quad \varphi_{1j} = (\omega_1 t - k_1 z_j), \tag{2}$$

$$\boldsymbol{E}_2(t, x_j) = E_{02} \cdot \exp\left[-\left(\frac{\varphi_{2j} - \delta\tau_2}{\omega_2 t_2}\right)^2\right] \cos\varphi_{2j} \cdot \boldsymbol{e}_y, \quad \varphi_{2j} = (\omega_2 t - k_2 x_j), \tag{3}$$

$$\boldsymbol{H}(t, z_j, x_j) = \boldsymbol{H}_1(t, z_j) + \boldsymbol{H}_2(t, x_j), \tag{4}$$

$$\boldsymbol{H}_1(t, z_j) = H_{01} \cdot \exp\left[-\left(\frac{\varphi_{1j} - \delta\tau_1}{\omega_1 t_1}\right)^2\right] \cos\varphi_{1j} \cdot \boldsymbol{e}_y, \tag{5}$$

$$\boldsymbol{H}_2(t, x_j) = H_{02} \cdot \exp\left[-\left(\frac{\varphi_{2j} - \delta\tau_2}{\omega_2 t_2}\right)^2\right] \cos\varphi_{2j} \cdot \boldsymbol{e}_z, \tag{6}$$

where $\varphi_{ij}$ are the phases of corresponding wave ($i = 1, 2$) and corresponding particle ($j = 1, 2$); $E_{0i}$ and $H_{0i}$ are the strength of electric and magnetic field in the pulse peak, respectively; $\delta\tau_i$ are the phase shifts of pulse peaks of the first and second waves; $t_i$ and $\omega_i$ are the pulse durations and frequency of the first and the second wave; $\boldsymbol{e}_x$, $\boldsymbol{e}_y$, $\boldsymbol{e}_z$ are unit vectors directed along the $x$, $y$ and $z$ axes.

It is well-known fact that in a frame of the dipole approximation ($k = 0$) and without taking into account terms of the order $v/c \ll 1$ the particle interaction with plane-wave field does not affect on the particle relative motion in center-of-mass system. Thereby, we consider particle motion in the laser field beyond to the dipole approximation and an accuracy of quantities of order $v/c \ll 1$ ($v$ is the relative velocity). Newton equations for motion of two identically charged particles with the mass $m$ and charge $e$ ($e = e_1 = e_2$) in the pulsed field of two mutually perpendicular laser waves (1) - (6) are determined by following expressions:

$$m\ddot{\boldsymbol{r}}_1^{(\mp)} = \mp |e| \left[ \boldsymbol{E}\left(t, z_1^{(\mp)}, x_1^{(\mp)}\right) + \frac{1}{c} \dot{\boldsymbol{r}}_1^{(\mp)} \times \boldsymbol{H}\left(t, z_1^{(\mp)}, x_1^{(\mp)}\right) \right] - \frac{e^2}{\left|\boldsymbol{r}_2^{(\mp)} - \boldsymbol{r}_1^{(\mp)}\right|^3} \left(\boldsymbol{r}_2^{(\mp)} - \boldsymbol{r}_1^{(\mp)}\right), \qquad (7)$$

$$m\ddot{\boldsymbol{r}}_2^{(\mp)} = \mp |e| \left[ \boldsymbol{E}\left(t, z_2^{(\mp)}, x_2^{(\mp)}\right) + \frac{1}{c} \dot{\boldsymbol{r}}_2^{(\mp)} \times \boldsymbol{H}\left(t, z_2^{(\mp)}, x_2^{(\mp)}\right) \right] + \frac{e^2}{\left|\boldsymbol{r}_2^{(\mp)} - \boldsymbol{r}_1^{(\mp)}\right|^3} \left(\boldsymbol{r}_2^{(\mp)} - \boldsymbol{r}_1^{(\mp)}\right), \qquad (8)$$

where $\boldsymbol{r}_1^{(\mp)}$ and $\boldsymbol{r}_2^{(\mp)}$ are electrons (sign of the charge is "–") or positrons (sign of the charge is "+") radius-vectors.

Hereafter, wave frequencies are same: $\omega_1 = \omega_2 = \omega$, $|\boldsymbol{k}_1| = |\boldsymbol{k}_2| = k = \omega/c = \lambdabar^{-1}$.

Subsequent consideration we carry out in the center-of-mass system:

$$\boldsymbol{r}^{(\mp)} = \boldsymbol{r}_2^{(\mp)} - \boldsymbol{r}_1^{(\mp)}. \qquad (9)$$

There are following equations for particle relative motion in the center-of-mass system:

$$m\ddot{\boldsymbol{r}}^{(\mp)} = \frac{2e^2}{\left|\boldsymbol{r}^{(\mp)}\right|^3} \boldsymbol{r}^{(\mp)} \mp |e| \left( \mathrm{M}_x^{(\mp)} \boldsymbol{e}_x + \mathrm{M}_y^{(\mp)} \boldsymbol{e}_y + \mathrm{M}_z^{(\mp)} \boldsymbol{e}_z \right), \qquad (10)$$

$$\begin{cases} \mathrm{M}_x^{(\mp)} = f_1' \left[ 2\sin(\omega t) \sin\left(k\frac{r_z^{(\mp)}}{2}\right) - \frac{1}{c} \dot{r}_z^{(\mp)} \cos(\omega t) \right] + \frac{1}{c} f_2' \dot{r}_y^{(\mp)} \cos(\omega t), \\ \mathrm{M}_y^{(\mp)} = f_2' \left[ 2\sin(\omega t) \sin\left(k\frac{r_x^{(\mp)}}{2}\right) - \frac{1}{c} \dot{r}_x^{(\mp)} \cos(\omega t) \right], \\ \mathrm{M}_z^{(\mp)} = \frac{1}{c} f_1' \dot{r}_x^{(\mp)} \cos(\omega t), \end{cases} \qquad (11)$$

$$f_1' = E_{01} \exp\left[ -\frac{(\omega t - \delta\tau_1)^2}{(\omega t_1)^2} \right], \quad f_2' = E_{02} \exp\left[ -\frac{(\omega t - \delta\tau_2)^2}{(\omega t_2)^2} \right]. \qquad (12)$$

Note that equations for relative motion (10) - (12) are written beyond to the dipole approximation ($k \neq 0$) and an accuracy of term of order $v/c \ll 1$. Note, small influence of external strong pulsed laser field on radius-vector of the center-of-mass motion was shown in the work [9]. Thereby, study of relative motion of electrons (positrons) makes sense to be done only.

Equations (10) - (11) have to be written in the dimensionless form:

$$\ddot{\boldsymbol{\xi}}^{(\mp)} = \boldsymbol{F}, \quad \boldsymbol{F} = \beta \cdot \frac{\boldsymbol{\xi}^{(\mp)}}{\left|\boldsymbol{\xi}^{(\mp)}\right|^3} \mp \left( N_x^{(\mp)} \boldsymbol{e}_x + N_y^{(\mp)} \boldsymbol{e}_y + N_z^{(\mp)} \boldsymbol{e}_z \right), \qquad (13)$$

$$\begin{cases} N_x^{(\mp)} = \eta_1 f_1 \left[ \sin(\tau) \sin\left(\frac{\xi_z^{(\mp)}}{2}\right) - \frac{\dot{\xi}_z^{(\mp)}}{2} \cos(\tau) \right] + \eta_2 f_2 \dot{\xi}_y^{(\mp)} \cos(\tau), \\ N_y^{(\mp)} = \eta_2 f_2 \left[ \sin(\tau) \sin\left(\frac{\xi_x^{(\mp)}}{2}\right) - \frac{\dot{\xi}_x^{(\mp)}}{2} \cos(\tau) \right], \\ N_z^{(\mp)} = \eta_1 f_1 \frac{\dot{\xi}_x^{(\mp)}}{2} \cos(\tau), \end{cases} \qquad (14)$$

where,

$$\xi^{(\mp)} = k r^{(\mp)} = r^{(\mp)}/\lambdabar, \quad \tau = \omega t, \quad \tau_{1,2} = \omega t_{1,2}, \qquad (15)$$

$$\eta_i = \frac{|e|E_{0i}}{\mu c \omega}, \quad \beta = \frac{e^2/\lambdabar}{\mu c^2}; \quad \mu = m/2, \quad f_i = \exp\left(-\frac{(\tau - \delta\tau_i)^2}{\tau_i^2}\right), \quad i = 1, 2. \qquad (16)$$

Here, $\xi^{(\mp)}$ is the radius-vector of the relative distance between electrons (positrons) in units of the wavelength, the parameters $\eta_{1,2}$ are numerically equal to ratio of an oscillation velocity of a particle in the peak of pulse of the first or second wave to the velocity of light $c$ (hereinafter, we consider parameters $\eta_{1,2}$ as oscillation velocities); the parameter $\beta$ is numerically equal to ratio of the energy of Coulomb interaction of particles with reduced mass $\mu$ at the wavelength to the particle rest energy. The pulse duration exceeds considerably the period of wave rapid oscillation ($\sim \omega^{-1}$) for a majority of modern pulsed lasers:

$$\tau_{1,2} \gg 1. \qquad (17)$$

Consequently, the relative distance between particles should be averaged over the period of wave rapid oscillation:

$$\overline{\xi}^{(\mp)} = \frac{1}{2\pi} \int_0^{2\pi} \xi^{(\mp)} \cdot d\tau. \qquad (18)$$

We emphasize that in the expression for the effective force $F$ (13) the first summand corresponds to the Coulomb repulsion of like-charged particles, and the second summand corresponds to the interaction of charged particles with the external laser field. It is important to note that for electrons and positrons this interaction has a different sign (see the sign before the second summand in expression (13)). Because of this, in the external laser field, the symmetry disappears in the interaction of particles of the same charge, but of different sign.

We note that expressions (13), (14) consider interaction with the Coulomb field and the pulsed-wave field strictly, and don't have the analytical solution. For subsequent analysis we will study all equations numerically.

Electrons (positrons) initial relative coordinates and velocities are the following:

$$\begin{aligned} &\xi_{x0}^{(\mp)} = 2, \quad \xi_{y0}^{(\mp)} = 0, \quad \xi_{z0}^{(\mp)} = 0, \\ &\dot{\xi}_{x0}^{(\mp)} = -1.7 \cdot 10^{-3}, \quad \dot{\xi}_{y0}^{(\mp)} = 0, \quad \dot{\xi}_{z0}^{(\mp)} = 0. \end{aligned} \qquad (19)$$

The interaction time is $\tau \in [-600 \div 600]$ and it was increased, if necessary for more clear results. Frequencies of waves are $\omega_1 = \omega_2 = 2\,\text{Ps}^{-1}$ ($\lambdabar = 0.15\,\mu\text{m}$), pulse durations are $\tau_1 = \tau_2 = 600$ ($t_1 = t_2 = 300\,\text{fs}$). Field intensities (oscillations velocities $\eta_1, \eta_2$) are varied. Phase shifts are vary within $\delta\tau_{1,2} \in [-600 \div 600]$ and step is $h = 50$. Initial conditions are the same as in [9]. That allows to estimate influence of phase shifts on relative motion of electrons and compare results. Note, in the work [9] the parameter of the phase shift of a pulse of a wave ($\delta\tau_{1,2} = 0$) was absent, and pulse peaks of both waves were in moment $\tau = 0$. Initial coordinates and velocities of electrons (positrons) are chosen so that at the point $\tau = 0$ particles were in maximum approach (the Coulomb force was maximum). In this work the pulse peaks of waves can have maximum at given moment of time ($\tau = \delta\tau_{1,2}$) and it's leads to significant change in the behavior of electron (positron) interaction. Numerical solving of equations for relative motion (13) results to several cases.

## 3. Anomalous repulsion

Here we consider the case when the oscillation velocity of the first wave is greater than the initial velocity of particles ($\eta_1 > \dot\xi_0^{\mp}$), and oscillation velocity of the second wave considerably exceeds the initial velocity ($\eta_2 \gg \dot\xi_0^{\mp}$). Let designate the final distance at which electrons and positrons scatter in the time moment $\tau_{(C)final}^{(\mp)} \equiv \tau_{(C)final} = 600$: without an external field as $\overline{\xi}_{(C)final}^{(\mp)} \equiv \overline{\xi}_{(C)final} = 2$; in the external field, when $\delta\tau_1 = 0$, $\delta\tau_2 = 0$ as $\overline{\xi}_{(0)final}^{(-)}$ for electrons and $\overline{\xi}_{(0)final}^{(+)}$ for positrons; in an external field, when $\delta\tau_1 \neq 0$, $\delta\tau_2 \neq 0$ as $\overline{\xi}_{final}^{(-)}$ for electrons and $\overline{\xi}_{final}^{(+)}$ for positrons.

Figures 2 and 3 show the average relative distances (in logarithmic units) between electrons (a) and positrons (b) against the interaction time $\tau$.

Figure 2 shows the average relative distances between electrons (see figure 2(a)) and positrons (see figure 2(b)) for the velocities of oscillations of two waves $\eta_1 = 3\times10^{-3}$, $\eta_2 = 6\times10^{-2}$, and for different values of the phase shifts of both waves (curves 1, 2, 3 in figure 2). For electrons and positrons, the same initial conditions, intensities, and phase shifts of the waves were chosen. It's seen from figure 2 that for phase shifts of waves having different signs (for the first wave the phase shift is negative, and for the second wave - positive), electrons and positrons abnormally strongly repulse. Moreover, it is possible to select such phase shifts of waves (see curves 3 in figure 2(a) and figure 2(b)), when the anomalous repulsion becomes maximum.

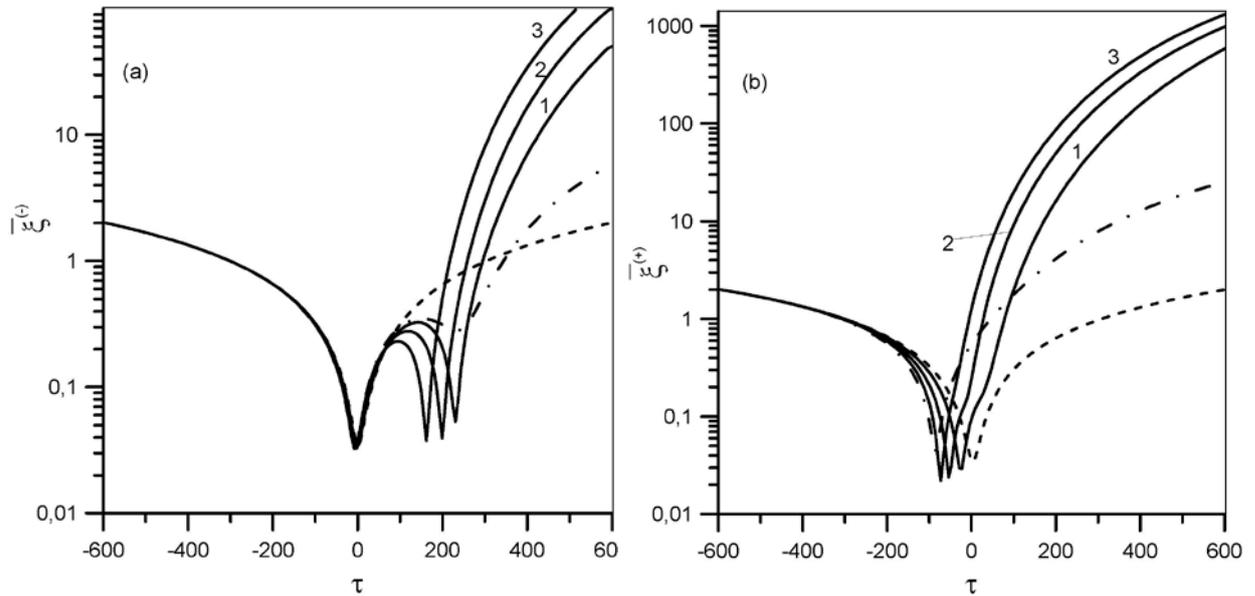

**Figure 2.** The averaged relative distance $\overline{\xi}^{(\mp)}$ of electrons (a) and positrons (b) against the interaction time $\tau$. The dashed line corresponds to the case of the absence of the external field. The dashed-dot line and solid lines correspond to oscillation velocities: $\eta_1 = 3\times10^{-3}$, $\eta_2 = 6\times10^{-2}$ (field intensities: $I_1 = 3.4\times10^{12}\,\text{W}/\text{cm}^2$, $I_2 = 1.3\times10^{15}\,\text{W}/\text{cm}^2$), phase shifts of pulse peaks: 1- $\delta\tau_1 = -450$, $\delta\tau_2 = 450$; 2- $\delta\tau_1 = -350$, $\delta\tau_2 = 300$; 3- $\delta\tau_1 = -550$, $\delta\tau_2 = 250$; the dashed-dot line - $\delta\tau_1 = 0$, $\delta\tau_2 = 0$.

Thus, for electrons at time $\tau = 600$ for phase shifts $\delta\tau_1 = -450$, $\delta\tau_2 = 450$, the ratio of the average relative distance in the laser field to the corresponding value without a field is $\overline{\xi}_{final}^{(-)} / \overline{\xi}_{(C)final} \approx 24$ (see curve 1 in figure 2(a)). At this, for positrons this ratio is $\overline{\xi}_{final}^{(+)} / \overline{\xi}_{(C)final} \approx 293$ (see curve 1 in figure 2(b)), i.e. an order of magnitude greater than the corresponding distance for electrons ($\overline{\xi}_{final}^{(+)} / \overline{\xi}_{final}^{(-)} \approx 12$). By analogy, for phase shifts $\delta\tau_1 = -350$, $\delta\tau_2 = 300$ for electrons $\overline{\xi}_{final}^{(-)} / \overline{\xi}_{(C)final} \approx 48$ (see curve 2 in figure 2(a)) and for positrons $\overline{\xi}_{final}^{(+)} / \overline{\xi}_{(C)final} \approx 506$ (see curve 2 in figure 2(b)). At this, the ratio of final positron and electron distances is $\overline{\xi}_{final}^{(+)} / \overline{\xi}_{final}^{(-)} \approx 10$. For phase shifts $\delta\tau_1 = -550$, $\delta\tau_2 = 250$ for electrons $\overline{\xi}_{final}^{(-)} / \overline{\xi}_{(C)final} \approx 81$ (see curve 3 in figure 2(a)) and for positrons $\overline{\xi}_{final}^{(+)} / \overline{\xi}_{(C)final} \approx 673$ (see curve 3 in figure 2(b)). The ratio of final positron and electron distances is $\overline{\xi}_{final}^{(+)} / \overline{\xi}_{final}^{(-)} \approx 8$. It is important to emphasize that with zero phase shifts, the effective repulsive force becomes significantly smaller (the corresponding averaged distances for electrons and positrons are an order of magnitude smaller).

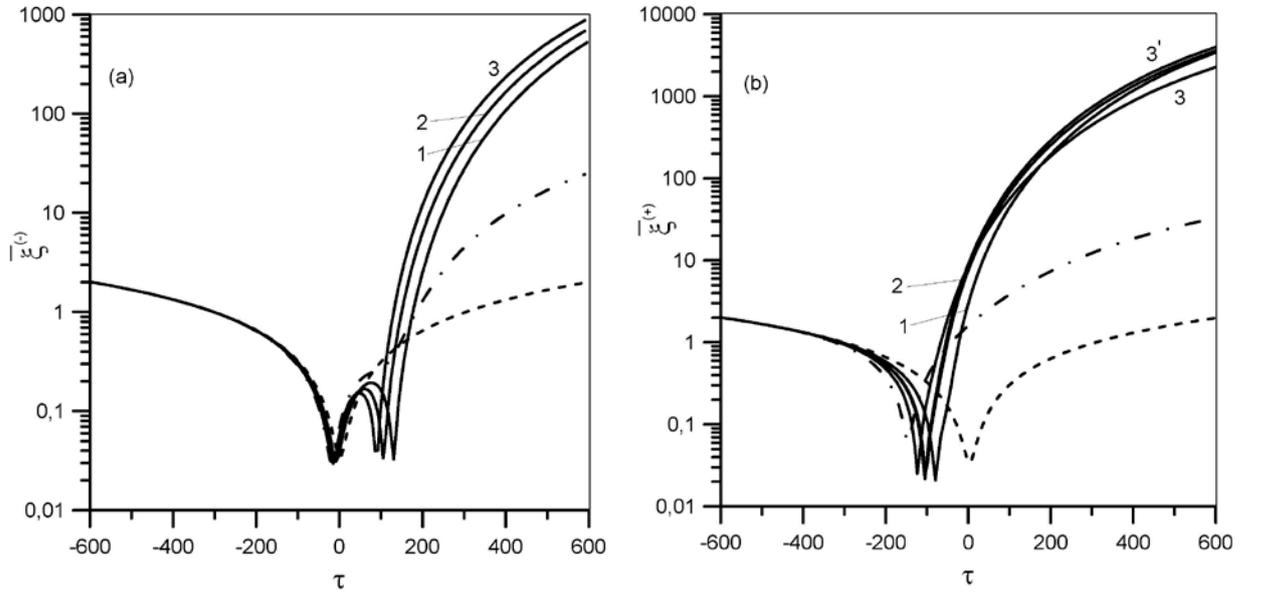

**Figure 3.** The averaged relative distance $\overline{\xi}^{(\mp)}$ of electrons (a) and positrons (b) against the interaction time $\tau$. The dashed line corresponds to case without external field. The dashed-dot line and solid lines correspond to oscillation velocities: $\eta_1 = 3 \times 10^{-3}$, $\eta_2 = 10^{-1}$ (the field intensities: $I_1 = 3.4 \times 10^{12} \text{ W}/\text{cm}^2$, $I_2 = 3.8 \times 10^{15} \text{ W}/\text{cm}^2$), the phase shifts of pulse peaks: 1- $\delta\tau_1 = -300, \delta\tau_2 = 350$; 2- $\delta\tau_1 = -400, \delta\tau_2 = 300$; 3- $\delta\tau_1 = -450, \delta\tau_2 = 250$; 3'- $\delta\tau_1 = -450, \delta\tau_2 = 300$; the dashed-dot line - $\delta\tau_1 = 0, \delta\tau_2 = 0$.

An increase in the rate of oscillations velocity of the second wave to a maximum value $\eta_2 = 10^{-1}$ leads to a significant increase in values of final distances, to which particles and antiparticles are scattered. In this case, for phase shifts $\delta\tau_1 = -300$, $\delta\tau_2 = 350$ and for electrons the ratio of the average relative distance in the laser field to the corresponding value without a field is $\overline{\xi}_{final}^{(-)} / \overline{\xi}_{(C)final} \approx 263$ (see curve 1 in figure 3(a)). For positrons, this ratio is $\overline{\xi}_{final}^{(+)} / \overline{\xi}_{(C)final} \approx 1670$ (see curve 1 in figure 3(b)). For phase shifts $\delta\tau_1 = -400$, $\delta\tau_2 = 300$ for electrons the ratio is $\overline{\xi}_{final}^{(-)} / \overline{\xi}_{(C)final} \approx 341$ (see curve 2 in figure

3(a)) and for positrons $\overline{\xi}^{(+)}_{final}/\overline{\xi}_{(C)final} \approx 1766$ (see curve 2 in figure 3(b)). For phase shifts $\delta\tau_1 = -450$, $\delta\tau_2 = 250$ for electrons the ratio is $\overline{\xi}^{(-)}_{final}/\overline{\xi}_{(C)final} \approx 437$ (see curve 3 in figure 3(a)) and for positrons $\overline{\xi}^{(+)}_{final}/\overline{\xi}_{(C)final} \approx 1110$ (see curve 3 in figure 3(b)). With not large increase in the phase shift of the second wave, the positrons scatter much further. So for phase shifts $\delta\tau_1 = -450$, $\delta\tau_2 = 300$ for positrons the ratio is $\overline{\xi}^{(+)}_{final}/\overline{\xi}_{(C)final} \approx 1903$ (see curve 3' in figure 3(b)).

Thus, for identical initial conditions and wave parameters, if the oscillation velocity of second wave is much larger than the initial velocity of particles and phase shifts of the waves have different signs, the effective interaction of two electrons and two positrons is appeared in their anomalous repulsion. In this case, the electrons can scatter over distances exceeding the corresponding values without a laser field up to two orders of magnitude, and positrons - up to three orders of magnitude. This indicates a significant asymmetry in the effective interaction of particles and antiparticles in strong pulsed laser fields. This asymmetry is associated with a positive sign of positrons charge, which leads to an increase in the effective repulsive force in comparison with electrons (see equations (13)).

## 4. The effective slowing-down of particles

In contrast to the previous section, we consider the case, when the oscillation velocity $\eta_1$ has to be close to the initial relative velocity $\eta_1 \approx \dot{\xi}_0^{\mp}$ and the oscillation velocity $\eta_2$ is greater an order of magnitude. Let designate the time at which the averaged relative distance between electrons or positrons is equal to $\overline{\xi}_{(C)final} = 2$: in the external field, when $\delta\tau_1 = 0$, $\delta\tau_2 = 0$ - $\tau^{(\mp)}_{(0)}$; in the external field, when $\delta\tau_1 \neq 0$, $\delta\tau_2 \neq 0$ - $\tau^{(\mp)}$.

In Figs. 4 and 5 show the average relative distances between electrons (see Figs. 4a and 5a) and positrons (see Figs. 4b and 5b) for oscillations velocities of two waves $\eta_1 = \dot{\xi}_0 = 1.7 \times 10^{-3}$, $\eta_2 = 3 \times 10^{-2}$, and for different values of phase shifts of both waves with same signs (for the first and second waves the phase shift is positive, see curves 1, 2 in figure 4 and figure 5).

For electrons and positrons, the same initial conditions, intensities, and phase shifts of the waves were chosen. It is seen from figure 4 that in the selected range of phase shifts $\delta\tau_1$, $\delta\tau_2 \in [300 \div 400]$, for electrons, we have the effect of attraction (see figure 4(a)), and for positrons the effect of repulsion (see figure 4(b)). At the same time, it is seen from figure 5 that for another set of values of phase shifts $\delta\tau_1$, $\delta\tau_2 \in [500 \div 600]$ we have the opposite figure 4 picture: positrons attract, and electrons repulse. Let's analyze in more detail figure 4 and figure 5.

Figure 4 shows that for phase shifts $\delta\tau_1 = 350$, $\delta\tau_2 = 400$ the time of electron scattering to the initial value of the distance ($\overline{\xi}^{(-)}_{final} = 2$) is increased in comparison with the case without an external field to $\tau^{(-)}/\tau_{(C)final} \approx 13.5$ (see curve 1 and the dashed line in figure 4(a)). In the external field when $\delta\tau_1 = 0$, $\delta\tau_2 = 0$ the time is increased to $\tau^{(-)}/\tau^{(-)}_{(0)} \approx 4$ (see curve 1 and the dashed-dot line in figure 4(a)). The effect is a bit weaker for phase shifts $\delta\tau_1 = 300$, $\delta\tau_2 = 300$ (see curve 2 in figure 4(a)). Under the same conditions, positrons effectively repulse and scatter over longer distances than without an external field $\overline{\xi}^{(+)}_{final}/\overline{\xi}_{(C)final} \approx 2.3$ (see curve 1 and the dashed line in figure 4(b)), but these distances are two times less than without the phase shifts $\overline{\xi}^{(+)}_{final}/\overline{\xi}^{(+)}_{(0)final} \approx 0.5$ (see curve 1 and the dashed-dot line in figure 4(b)). Note that for curve 2 in figure 4(b) we have similar results $\overline{\xi}^{(+)}_{final}/\overline{\xi}_{(C)final} \approx 3.3$ (see curve 2 and the dashed line in figure 4(b)) and $\overline{\xi}^{(+)}_{final}/\overline{\xi}^{(+)}_{(0)final} \approx 0.7$ (see curve 2 and the dashed-dot line in figure 4(b)).

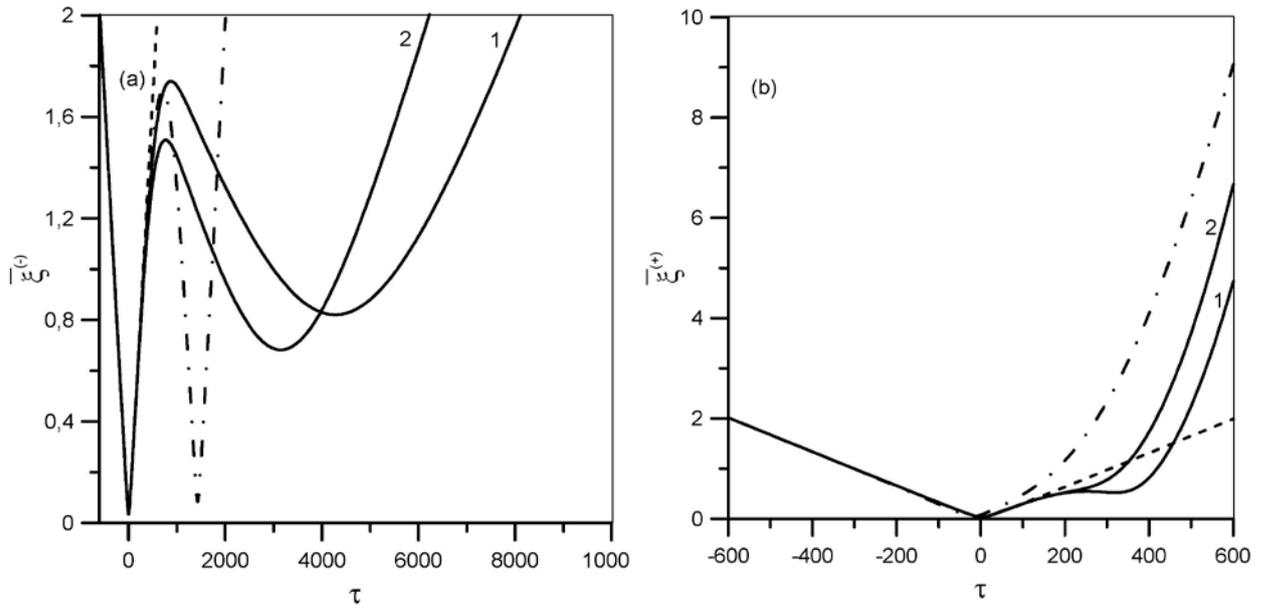

**Figure 4.** The averaged relative distance $\bar{\xi}^{(\mp)}$ of electrons (a) and positrons (b) against the interaction time $\tau$. The dashed line corresponds to case without external field. The dashed-dot line and solid lines correspond to the oscillation velocity: $\eta_1 = \dot{\xi}_0 = 1.7 \times 10^{-3}$, $\eta_2 = 3 \times 10^{-2}$ (the field intensity: $I_1 = 1.1 \times 10^{12}\ \mathrm{W/cm^2}$, $I_2 = 3.4 \times 10^{14}\ \mathrm{W/cm^2}$), phase shifts of pulse peaks: 1- $\delta\tau_1 = 350$, $\delta\tau_2 = 400$, 2- $\delta\tau_1 = 300$, $\delta\tau_2 = 300$; the dashed-dot line - $\delta\tau_1 = 0$, $\delta\tau_2 = 0$.

Figure 5 shows that for phase shifts $\delta\tau_1 = 550$, $\delta\tau_2 = 500$ electrons repulse and scatter over longer distances than without an external field: $\bar{\xi}^{(-)}_{final}/\bar{\xi}_{(C)final} \approx 1.8$ (see curve 1 and the dashed line in figure 5(a)) and in an external field without phase shifts: $\bar{\xi}^{(-)}_{final}/\bar{\xi}^{(-)}_{(0)final} \approx 4.3$ (see curve 1 and the dashed-dot line in figure 5(a)). For phase shifts $\delta\tau_1 = 550$, $\delta\tau_2 = 600$ we have similar results: $\bar{\xi}^{(-)}_{final}/\bar{\xi}_{(C)final} \approx 1.7$ (see curve 2 and the dashed line in figure 5(a)) and in an external field without phase shifts: $\bar{\xi}^{(-)}_{final}/\bar{\xi}^{(-)}_{(0)final} \approx 4$ (see curve 2 and the dashed-dot line in figure 5(a)). Unlike electrons, with the same phase shifts, positrons are effectively attracted and scattered to the initial distance in a longer time than without an external field: $\tau^{(+)}/\tau_{(C)final} \approx 1.3$ (see curve 1 and the dashed line in figure 5(b)) and even more slowly than in an external field without phase shifts: $\tau^{(+)}/\tau^{(+)}_{(0)} \approx 2.9$ (see curve 1 and the dashed-dot line in figure 5(b)).

With increasing intensity of the second wave to maximum values $\eta_2 = 10^{-1}$ and same phase shifts as in figure 5 ($\delta\tau_1$, $\delta\tau_2 \in [500 \div 600]$), the behavior of electrons and positrons becomes similar (repulsion takes place) (see Figures 6(a), (b)).

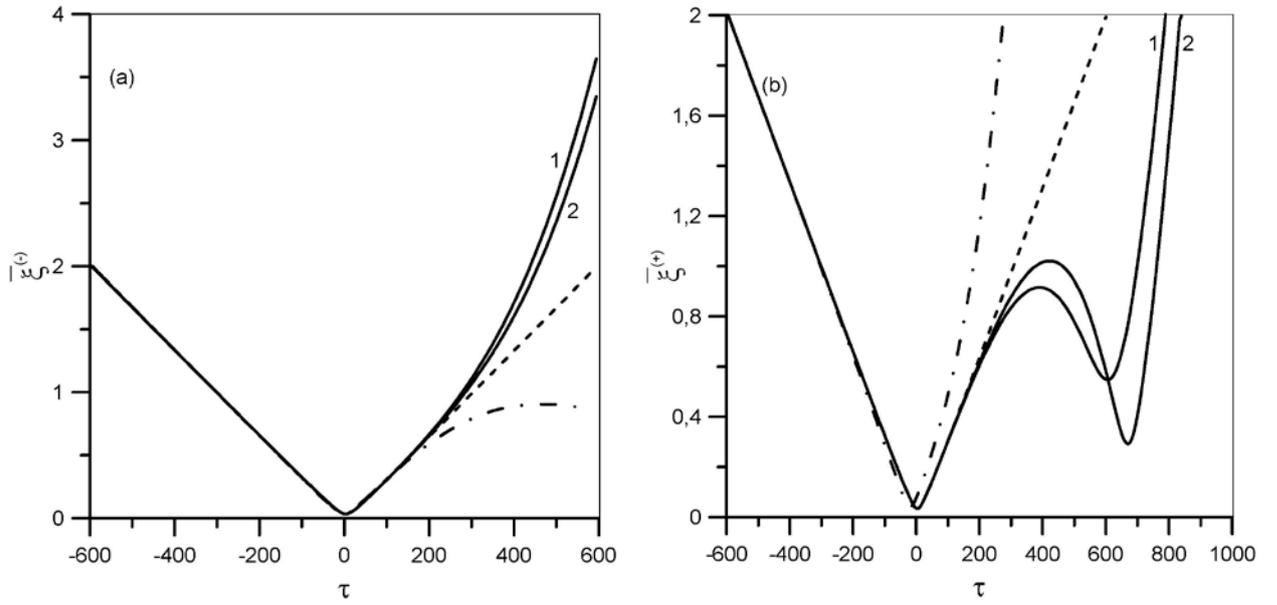

**Figure 5.** The averaged relative distance $\bar{\xi}^{(\mp)}$ of electrons (a) and positrons (b) against the interaction time $\tau$. The dashed line corresponds to case without external field. The dashed-dot line and solid lines correspond to the oscillation velocity: $\eta_1 = \dot{\xi}_0 = 1.7 \times 10^{-3}$, $\eta_2 = 3 \times 10^{-2}$ (the field intensity: $I_1 = 1.1 \times 10^{12}\,\text{W}/\text{cm}^2$, $I_2 = 3.4 \times 10^{14}\,\text{W}/\text{cm}^2$), phase shifts of pulse peaks: 1- $\delta\tau_1 = 550$, $\delta\tau_2 = 500$, 2- $\delta\tau_1 = 550$, $\delta\tau_2 = 600$; the dashed-dot line - $\delta\tau_1 = 0$, $\delta\tau_2 = 0$.

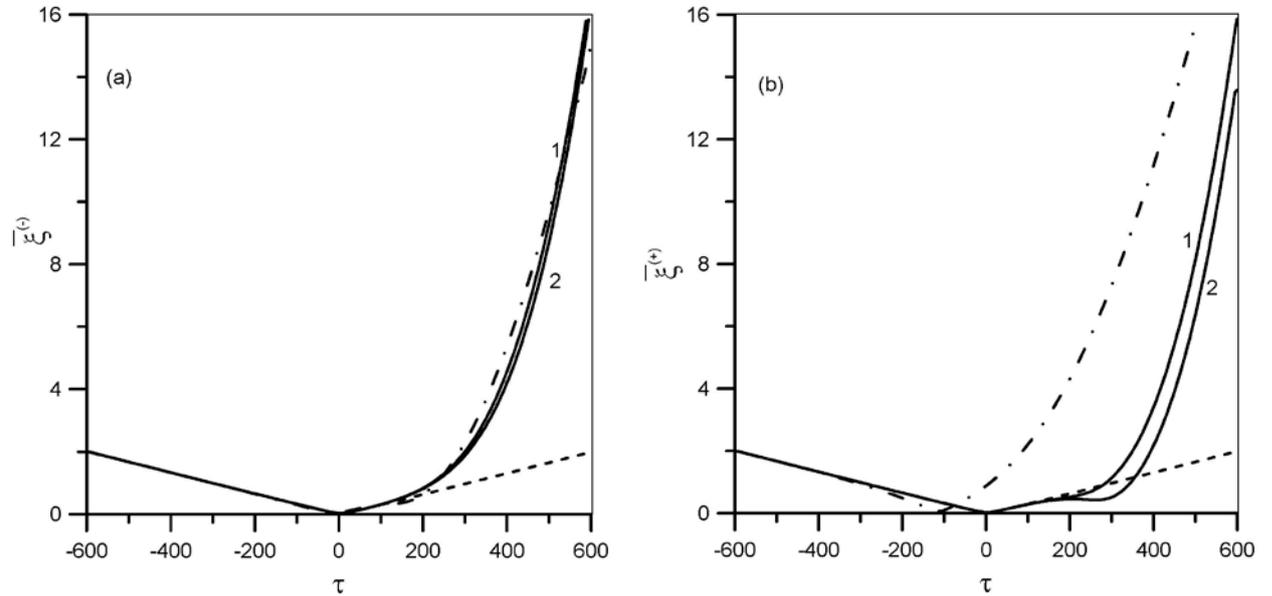

**Figure 6.** The averaged relative distance $\bar{\xi}^{(\mp)}$ of electrons (a) and positrons (b) against the interaction time $\tau$. The dashed line corresponds to case without external field. The dashed-dot line and solid lines correspond to the oscillation velocity: $\eta_1 = \dot{\xi}_0 = 1.7 \times 10^{-3}$, $\eta_2 = 10^{-1}$ (the field intensity: $I_1 = 1.1 \times 10^{12}\,\text{W}/\text{cm}^2$, $I_2 = 3.8 \times 10^{15}\,\text{W}/\text{cm}^2$), phase shifts of pulse peaks: 1- $\delta\tau_1 = 550$, $\delta\tau_2 = 500$, 2- $\delta\tau_1 = 550$, $\delta\tau_2 = 600$; the dashed-dot line - $\delta\tau_1 = 0$, $\delta\tau_2 = 0$.

## 5. Conclusion

The study of the interaction of two electrons and two positrons in strong laser fields has shown a significant asymmetry in the effective interaction of particles and antiparticles. This asymmetry essentially depends on the magnitude and sign of phase shifts of waves relative to each other and intensities of both waves.

1. If the phase shifts of the waves have different signs (for the first wave the phase shift is negative, and for the second wave - positive), electrons and positrons anomalous strongly repulse. Moreover, at the maximum intensity of the second wave, it is possible to select phase shifts of waves when the anomalous repulsion of particles becomes maximal, exceeding the averaged scatter distances of particles without a field: for electrons by two orders of magnitude, and for positrons by three orders of magnitude (see figure 2 and figure 3).
2. If the phase shifts of both waves have the same sign (for the first and second waves the phase shift is positive), then the interaction of the particles varies significantly. If the velocities of the wave oscillations do not differ much from each other ($\eta_1 = \dot{\xi}_0 = 1.7 \times 10^{-3}$, $\eta_2 = 3 \times 10^{-2}$) and the values of phase shifts are in the range $\delta\tau_1$, $\delta\tau_2 \in [300 \div 400]$, then for electrons there is an attractive effect (see figure 4(a)), and for positrons - the effect of repulsion (see figure 4(b)). In this case, for another set of values of phase shifts $\delta\tau_1$, $\delta\tau_2 \in [500 \div 600]$ we have the opposite situation: positrons attract, and electrons repulse (see figure 5).
3. In the range of phase shifts $\delta\tau_1$, $\delta\tau_2 \in [500 \div 600]$ and at the maximum intensity of the second wave ($\eta_2 = 0.1$) the effective interaction of electrons and positrons occurs in a similar form (see figure 6).

The obtained results can be used for experiments in the framework of modern research projects, where the sources of pulsed laser radiation are used (SLAC, FAIR) [10-12].